\documentclass[journal,twoside,web]{ieeecolor}
\usepackage{tmi}
\usepackage{cite}
\usepackage{amsmath,amssymb,amsfonts}
\usepackage{algorithm}
\usepackage{algorithmic}
\usepackage{graphicx}
\usepackage{textcomp}
\usepackage{hyperref}
\usepackage{booktabs} 
\usepackage{multirow}
\usepackage{etoolbox}
\usepackage{rotating}

\makeatletter
\@ifundefined{color@begingroup}%
{\let\color@begingroup\relax
\let\color@endgroup\relax}{}%
\def\fix@ieeecolor@hbox#1{%
\hbox{\color@begingroup#1\color@endgroup}}
\patchcmd\@makecaption{\hbox}{\fix@ieeecolor@hbox}{}{\FAILED}
\patchcmd\@makecaption{\hbox}{\fix@ieeecolor@hbox}{}{\FAILED}
\def\BibTeX{{\rm B\kern-.05em{\sc i\kern-.025em b}\kern-.08em
    T\kern-.1667em\lower.7ex\hbox{E}\kern-.125emX}}
\begin{document}
\title{Exploiting Manifold Structured Data Priors for Improved MR Fingerprinting Reconstruction}
\author{Peng Li, \IEEEmembership{Student Member, IEEE}, Yuping Ji, \IEEEmembership{Student Member, IEEE}, Yue Hu, \IEEEmembership{Member, IEEE}
\thanks{This work is supported by the Natural Science Foundation of Heilongjiang YQ2021F005 and China NSFC 61871159.}
\thanks{Y. Hu, P. Li, and Yuping Ji are with the School of Electronics and Information Engineering, Harbin Institute of Technology, Harbin, China (e-mail: huyue@hit.edu.cn).}
\thanks{Our code will be available at \url{https://github.com/bigponglee/MS-LLR-MRF}.}
}

\maketitle

\begin{abstract}
  Estimating tissue parameter maps with high accuracy and precision from highly undersampled measurements presents one of the major challenges in MR fingerprinting (MRF). Many existing works project the recovered voxel fingerprints onto the Bloch manifold to improve reconstruction performance. However, little research focuses on exploiting the latent manifold structure priors among fingerprints. To fill this gap, we propose a novel MRF reconstruction framework based on manifold structured data priors. Since it is difficult to directly estimate the fingerprint manifold structure, we model the tissue parameters as points on a low-dimensional parameter manifold. We reveal that the fingerprint manifold shares the same intrinsic topology as the parameter manifold, although being embedded in different Euclidean spaces. To exploit the non-linear and non-local redundancies in MRF data, we divide the MRF data into spatial patches, and the similarity measurement among data patches can be accurately obtained using the Euclidean distance between the corresponding patches in the parameter manifold. The measured similarity is then used to construct the graph Laplacian operator, which represents the fingerprint manifold structure. Thus, the fingerprint manifold structure is introduced in the reconstruction framework by using the low-dimensional parameter manifold. Additionally, we incorporate the locally low-rank prior in the reconstruction framework to further utilize the local correlations within each patch for improved reconstruction performance. We also adopt a GPU-accelerated NUFFT library to accelerate reconstruction in non-Cartesian sampling scenarios. Experimental results demonstrate that our method can achieve significantly improved reconstruction performance with reduced computational time over the state-of-the-art methods.
\end{abstract}

\vspace{-5pt}
\section{Introduction}
\label{sec:introduction}
\IEEEPARstart{M}{agnetic} resonance fingerprinting (MRF) is a promising quantitative MRI method proposed by Ma {\em et al}. \cite{ma2013magnetic}\cite{chen2022technical}, which enables simultaneous imaging of multiple tissue parameters, including spin-lattice relaxation time (${\rm T}_1$), spin-spin relaxation time (${\rm T}_2$), etc. MRF utilizes a variable schedule of radiofrequency excitations and delays to induce unique signal evolutions from different tissues, which are termed fingerprints. The multiple quantization parameters are then obtained by mapping acquired fingerprints to a precomputed dictionary that contains theoretical signal evolutions of a set of tissues using pattern-matching algorithms. However, rapid acquisition schemes are widely used in MRF to accelerate data acquisition, leading to aliasing artifacts in the recovered fingerprints \cite{tippareddy2021magnetic}. Fingerprints contaminated with aliasing artifacts and noise will subsequently decrease the accuracy of estimated tissue parameters. \footnote{This work has been submitted to the IEEE Transactions on Medical Imaging.}
\footnote{© 20XX IEEE. Personal use of this material is permitted. Permission from IEEE must be obtained for all other uses, in any current or future media, including reprinting/republishing this material for advertising or promotional purposes, creating new collective works, for resale or redistribution to servers or lists, or reuse of any copyrighted component of this work in other works.}

To improve the accuracy of the reconstructed parameter maps, many reconstruction methods have been proposed to overcome undersampling artifacts. Davis {\em et al}. \cite{davies2014compressed} proposed to apply Bloch response manifold projection in a compressed sensing framework (BLIP) to improve MRF reconstruction. Zhao {\em et al}. \cite{zhao2016maximum} proposed a maximum likelihood formalism (MBIR) to estimate multiple parameter maps directly from highly undersampled data. While the aforementioned algorithms have shown improved performance compared with the original MRF method, they do not take full advantage of the temporal and spatial correlation of MRF data. 

The latest research focuses more on using the temporal and spatial prior constraints of MRF data to further improve the performance of reconstruction algorithms. Mazor {\em et al}. \cite{mazor2018low} developed a subspace-constrained low-rank projection method (FLOR), based on the fact that the MRF signal can be sparsely represented in the generated dictionary domain. Zhao {\em et al}. \cite{zhao2018improved} proposed a constrained imaging method based on low-rank and subspace modeling to improve the accuracy and speed of MRF. Then, they \cite{zhao2020further} extended this method by introducing a low-rank tensor model, which can mitigate the information loss caused by the matrix preprocessing step in the low-rank reconstruction method. Cruz {\em et al}. \cite{lima2019sparsity} developed a sparse and locally low-rank regularized reconstruction method, enabling shorter scan times and increased spatial resolution. In our previous work, we also proposed a structured low-rank matrix completion and subspace projection framework (SL-SP) \cite{hu2021high} to recover MRF data from its highly undersampled measurements, resulting in improved reconstruction performance. Nagtegaal {\em et al}. \cite{nagtegaal2023multicomponent} proposed to obtain multicomponent parameter maps directly from MRF $k$-space measurements by using joint-sparsity and low-rank constraints. Leveraging prior data from MRF has great potential to improve reconstruction performance. However, these methods are limited in their ability to treat MRF data solely as a low-rank matrix or utilize only local prior information. A significant research gap remains in harnessing more promising non-local and non-linear priors in MRF.

The above methods improved the accuracy of the reconstructed parameter maps by enhancing the quality of the reconstructed MRF data. Another solution is to improve the speed and accuracy of the pattern-matching method to reconstruct high-precision parameter maps. McGivney {\em et~al.} \cite{mcgivney2014svd} proposed a singular value decomposition (SVD)-based method to project the dictionary and voxel fingerprints into a low-dimensional subspace along the time domain and perform pattern matching in the low-dimensional subspace, resulting in reduced matching time and computational overhead. Yang {\em et~al.} \cite{yang2018low} further proposed to use randomized SVD to directly estimate the low-dimensional dictionary subspace, which can simultaneously reduce the computational time and the memory demand of the dictionary. However, these methods offer a limited acceleration in parameter reconstruction and may introduce errors in subsequent processes due to information loss associated with SVD. 

To address this problem, many deep learning-based methods have been introduced in MRF. Cohen {\em et al}. \cite{cohen2018mr} proposed a 4-layer fully connected neural network to perform signal-to-parameter mapping, replacing the memory-intensive dictionary and time-consuming dictionary matching. Oksuz {\em et al}. \cite{oksuz2019magnetic} proposed a recurrent neural network to perform MRF map reconstruction, exploiting the time-dependent information of tissue fingerprints. Fang {\em et al}. \cite{fang2019deep} proposed a two-step deep learning model (SCQ) to learn the mapping from the signals to the tissue parameters, enabling accurate parametric reconstructions under quadruple accelerated acquisition. Soyak {\em et al}. \cite{soyak2021channel} proposed a neural network consisting of a channel-wise attention module and a fully convolutional network, and the strategy of overlapping patches for patch-level multi-parameter estimation was adopted to effectively reduce the error of parameter reconstruction. However, the need for large training datasets limits the development of deep learning techniques in the field of MRF, and the generalizability of these existing studies remains to be verified.

Recently, manifold models have been explored as competitive alternatives for MRI reconstruction \cite{tenenbaum2000global, van2009dimensionality,aljabar2012manifold, boumal2023introduction}. Manifold-based methods treat image frames or measurements as points on a low-dimensional manifold embedded in a high-dimensional space, utilizing non-linear and non-local manifold structured data priors to improve reconstruction performance. Poddar {\em et al}. \cite{poddar2015dynamic} proposed a dynamic MRI reconstruction method (SToRM) by modeling image frames as points on a smooth, low-dimensional manifold in high-dimensional space, exploiting the non-linear and non-local redundancies in the data for improved reconstruction performance. In addition, the SToRM method was extended to bandlimited image manifold \cite{poddar2019manifold} and navigator-free sub-patches manifold \cite{mohsin2019free} in the subsequent studies. Nakarmi {\em et al}. \cite{nakarmi2017kernel} utilized kernel principal component analysis to learn the underlying manifold described by the principal components of the feature space, and enforced such structure through low-rank constraint in feature space, accelerating dynamic MRI reconstruction. Slavakis {\em et al}. \cite{slavakis2022kernel} proposed a non-parametric approximation framework for imputation-by-regression on data with missing entries, assuming that the data features lie close to a smooth manifold in a reproducing kernel Hilbert space. Djebra {\em et al}. \cite{djebra2022manifold} proposed the LTSA method for accelerated dynamic MRI, which aligns the local coordinates of the smooth manifold learned by linear subspace approximation with the global coordinates. Manifold structured data priors have shown great potential in efficiently exploiting the non-local and non-linear structural information of high-dimensional data. However, it is still challenging to accurately estimate the latent manifold from the acquired high-dimensional data, particularly in the presence of noise and undersampling. 

Although many studies \cite{zhao2018improved}\cite{mazor2018low}\cite{arberet2021parallel} have attempted to project the recovered fingerprints onto the Bloch manifold to improve reconstruction performance, they do not utilize the latent manifold structure priors in MRF. To fill this gap, in this paper, we propose a novel MRF reconstruction framework that utilizes both manifold structured data priors and locally low-rank constraints, termed MS-LLR. Inspired by the prior knowledge that fingerprints belong to the Bloch manifold, we propose to exploit the manifold structure of the high-dimensional MRF data to regularize the reconstruction problem. However, since it is difficult to directly estimate the fingerprint manifold structure, we further model the tissue parameters as points on a low-dimensional parameter manifold. In addition, we reveal that the fingerprint manifold share the same intrinsic topology as the parameter manifold, although being embedded in different Euclidean spaces. However, fingerprints contaminated with aliasing artifacts and noise will decrease the accuracy of estimated tissue parameters, thereby reducing the accuracy of the estimated manifold structure. To overcome this problem, we propose to enforce the manifold structured data priors in a spatially patch-wise manner, more accurately exploiting the non-linear and non-local redundancies in MRF data. We divide the MRF data into spatial patches and measure the similarity among data patches by using the Euclidean distance between the corresponding patches in the parameter manifold. The measured similarity is then used to construct the graph Laplacian operator, which represents the fingerprint manifold structure. Thus, the manifold structured data prior is introduced in the reconstruction framework by using the low-dimensional parameter manifold. Although the manifold structured data priors can efficiently utilize the non-local and non-linear structural features of high-dimensional MRF data, they may not be able to capture local correlations within each data patch. To overcome this limitation, we incorporate the locally low-rank regularization to further improve the reconstruction performance. We also adopt a GPU-accelerated NUFFT library to accelerate reconstruction in non-Cartesian sampling scenarios. Experimental results demonstrate that our proposed method can achieve significantly improved reconstruction performance with greatly reduced computational time over the state-of-the-art methods.

\vspace{-5pt}
\section{Background}
\label{sec:background}
\subsection{Manifold Regularization}
\label{sec:manifold}
\begin{figure}[htbp]   
  \vspace{-10pt}
  \centering   
  \includegraphics[width=180pt]{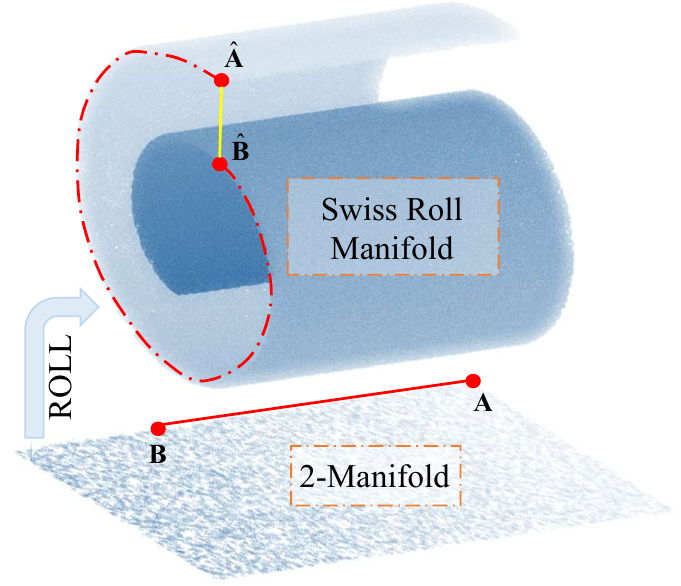}   
  \vspace{-8pt}
  \caption{Illustration of the well-known Swiss roll manifold: 2-manifold embedded in 3D space. For two arbitrary points $\hat{\mathbf{A}}$ and $\hat{\mathbf{B}}$ on the Swiss roll manifold, their Euclidean distance in the high-dimensional Euclidean space (yellow solid line) may not accurately reflect their intrinsic similarity. The intrinsic similarity can be accurately measured by the geodesic distance (red dotted line) between two points on the manifold or the Euclidean distance between the corresponding points on the lower dimensional 2-manifold (red solid line). 
  }   
  \label{swissroll}   
  \vspace{-20pt}
\end{figure}

The manifold hypothesis states that acquired real-world data lie on low-dimensional manifolds embedded within the high-dimensional space \cite{tenenbaum2000global, aljabar2012manifold, boumal2023introduction}. To further elaborate, the dimension of a manifold is typically lower than that of the ambient space within which it is embedded. Note that to distinguish the dimensions of manifolds from those of Euclidean space, we abbreviate a $n$-dimensional manifold as a $n$-manifold in the following text.  A classic illustration of a manifold is the Swiss roll manifold, as shown in Fig.\ref{swissroll}. It is a 2-manifold embedded in the 3D Euclidean space, where a plane 2-manifold is ``rolled'' into a cylinder shape. This reveals different representations of the same manifold embedded in different Euclidean spaces. The Euclidean distance between two points in the high-dimensional space may not accurately reflect their intrinsic similarity \cite{tenenbaum2000global}. For instance, two arbitrary points $\hat{\mathbf{A}}$ and $\hat{\mathbf{B}}$ on the Swiss roll manifold may have a smaller Euclidean distance (yellow solid line in Fig.\ref{swissroll}) than their intrinsic similarity (red dotted line). The intrinsic similarity can be measured by the geodesic distance between two points on the manifold or the Euclidean distance between the corresponding points ($\mathbf{A}$ and $\mathbf{B}$) on the lower dimensional 2-manifold (red solid line). The former requires an accurate estimation of the latent topology of the manifold, whereas the latter requires a proper dimension reduction to recover the undistorted 2-manifold from its high-dimensional embedding. However, it is challenging to accurately estimate the latent manifold structure and its dimensionality, as well as perform proper dimension reduction of high-dimensional acquired data, particularly in the presence of noise and undersampling.
\subsection{MRF Reconstruction Model}
Rapid acquisition schemes are widely used in MRF to accelerate data acquisition \cite{tippareddy2021magnetic}, which can be modeled as:
\begin{equation}
  \label{data_ac}
  \mathbf b={\cal A} {\cal X} +\mathbf n
\end{equation}
where ${\mathbf b} \in {\mathbb C}^{N_c \times N_s \times L}$ is the acquired $k$-space measurements with $N_c$ coils, $N_s$ is the number of $k$-space samples collected by each coil in each frame, and $L$ is the number of time frames. ${\cal X} \in {\mathbb C}^{N_x \times N_y \times L}$ denotes the distortion-free MRF data, where $N_x$ and $N_y$ are the image dimensions of each image frame. $\mathbf n \in {\mathbb C}^{N_c \times N_s \times L}$ is the noise matrix of all coils. ${\cal A} = {\mathbf F}_u\mathbf C$ denotes a linear operator that considers the coil sensitivities $\mathbf C$ and the undersampled Fourier transform $\mathbf F_u$. 

According to the MRF imaging mechanism \cite{davies2014compressed}, the magnetization response at any voxel of ${\cal X}$ can be written as a parametric nonlinear mapping as:
\begin{equation} 
  \label{mrf}
	{\cal X}_{i,j,:}=\rho_{i,j} \mathbf{B}(\eta_{i,j}; \theta)
\end{equation}
where $\rho_{i,j} \in {\mathbb R}_+ $ represents the proton density ($\rm PD$) of the corresponding voxel. $\eta_{i, j} =[{\rm T}_1, {\rm T}_2, \cdots]$ represents a row vector composed of different parameters. Note that three main parameters ${\rm T}_1$, ${\rm T}_2$ and ${\rm PD}$ are considered in this study, while the model can be easily generalized to include more parameters. We model the 3-dimensional parameter manifold as ${\cal M} \in {\mathbb{R}}^3$ to provide the set of feasible values for tissue parameters $\rho$ and $\eta$. $\theta$ is the parameter vector of the excitation pulse with length $L$, including the repetition time (TR), echo time (TE), and the flip angle (FA). $\mathbf{B}(\cdot ): {\mathbb{R}}^2\rightarrow {\mathbb{C}^{L}}$ denotes a smoothing mapping induced by the Bloch equation dynamic. 

The dictionary in MRF can be precomputed offline using the Bloch equation, which is modeled as:
\begin{equation} 
  \label{D}
	\mathbf{D}=\{\mathbf{d}_{m,:}\},\qquad \mathbf{d}_{m,:}=\mathbf{B}(\eta_m ,\theta) \in {\mathbb{C}^{1\times L}} 
\end{equation}
where \{$\eta_{m},  m=1,\cdots, M$\} represents a set of discrete parameters taken from the parameter manifold ${\cal M}$. $\mathbf{D}$ denotes the constructed dictionary whose entries represent theoretical response signal evolutions for a set of possible tissues. Note that the proton density of each entry $\mathbf{d}_{m,:}$ in the dictionary is set to 1. Meanwhile, a parameter look-up table ($\mathbf{LUT}$) can be obtained to record tissue parameters for the corresponding dictionary entry. The dictionary $\mathbf{D}$ is essentially a discretized approximation to the Bloch response manifold ${\cal B}$ \cite{davies2014compressed}.

Multiple parameters can be estimated by matching the reconstructed MRF data $\hat{\cal X}$ with the dictionary entries \cite{ma2013magnetic}, which can be expressed as:
\begin{equation} 
  \label{match}
  \hat{\mathbf{M}} =\varPhi_{\mathbf{D}}(\hat{\cal X})
\end{equation}
where $\hat{\mathbf{M}}$ denotes the estimated multiple parameter maps. The matching operator $\varPhi_{\mathbf{D}}$ reconstruct parameter maps from the MRF data $\hat{\cal X}$ using the precomputed dictionary $\mathbf{D}$ and $\mathbf{LUT}$, which can be defined as:
\begin{equation} 
  \label{matchf}
  \varPhi_{\mathbf{D}}({\cal X}):\left\{
  \begin{aligned}
    &\ k^{i,j}=\mathop{\arg\max}_{k}\frac{|\langle \mathbf{d}_{k,:}, {\cal X}_{i,j,:}\rangle |}{\lVert \mathbf{d}_{k,:} \rVert_2^2}  \\
    &\ \hat{\eta}_{i,j}=\mathbf{LUT}[k^{i,j}]\\
    &\  \hat{\rho}_{i,j}=\max\left\{ \frac{real\langle \mathbf{d}_{k^{i,j},:}, {\cal X}_{i,j,:}\rangle }{\lVert \mathbf{d}_{k^{i,j},:} \rVert_2^2},0\right\}
  \end{aligned}
  \right.
\end{equation}
where $k^{i,j}$ represents the index of the best matching entry in the dictionary with the corresponding signal, $\hat{\eta}_{i,j}$ and $\hat{\rho}_{i,j}$ denote estimated parameters vector and estimated proton density, respectively.
\vspace{-5pt}
\section{Methods}
\label{sec:methods}
\subsection{Manifold structured data Regularization}
\label{sec:manifold_reg}
Once the excitation pulse parameters $\theta$ are determined, the Bloch manifold $\cal B$ can be completely characterized using the tissue parameters ($\{{\rm T}_1, {\rm T}_2 \}$) and the Bloch equation dynamic $\mathbf{B}(\cdot )$. Thus, the Bloch manifold $\cal B$ can be considered as a 2-manifold embedded in an $L$ dimensional space through the Bloch equation dynamic. Furthermore, we use $\cal S$ to denote the fingerprint manifold, which is the cone \cite{davies2014compressed}\cite{cannon1979shrinking} on the Bloch response manifold $\cal B$. According to (\ref{mrf}), the magnetization response at any voxel lies on the fingerprint manifold $\cal S$, which can be regarded as a $3$-manifold embedded in the $L$-dimensional space. To clarify, we list the important symbols used in the paper in Table.\ref{symbols}.

\begin{table}[htbp]
  \vspace{-12pt}
  \centering
  \caption{Several important symbols and their descriptions.}
  \resizebox{170pt}{!}{
    \begin{tabular}{ll}
    \toprule
    Symbol & Description  \\
    \midrule
    $\mathbf{B}$ & Bloch equation dynamic operator \\
    $\cal B$ & Bloch response manifold \\
    $\mathbf{M}$ & Multiple parameter maps \\
    ${\cal M}$ & Tissue parameter manifold \\
    $\cal S$ & Fingerprint manifold \\
    \bottomrule
    \end{tabular}}%
  \label{symbols}%
  \vspace{-7pt}
\end{table}%

As proved in \cite{dong2019quantitative}, if two voxel fingerprints are sufficiently close, the corresponding tissue parameters also exhibit analogous proximity, i.e.:
\begin{equation}
  \label{hold}
  \lVert \rho_i\mathbf{B}(\eta_i; \theta) - \rho_j\mathbf{B}(\eta_j; \theta) \rVert_F^2 \leq \delta \Leftrightarrow \lVert \eta_i - \eta_j \rVert_F^2 \leq C\delta
\end{equation}
where $\delta>0$ is a small constant, and the constant $C>0$ is independent of $\delta$. $\lVert \cdot \rVert_F$ denotes the Frobenius norm. The above property indicates that the Bloch equation dynamic $\mathbf{B}(\cdot )$ provides a stable non-linear mapping between the tissue fingerprint and the corresponding parameters. Thus, in the manifold learning framework, the distance between any two points on the fingerprint manifold $\cal S$ can be well preserved in the corresponding points on the parameter manifold $\cal M$. In addition, based on MRF physics, any point on the fingerprint manifold $\cal S$ corresponds to a unique point on the parameter manifold $\cal M$. These two aspects guarantee that the fingerprint manifold $\cal S$ and the parameter manifold $\cal M$ share the same intrinsic topology, although being embedded in different Euclidean spaces.

The manifold regularization has been widely used in machine learning applications \cite{van2009dimensionality}. This scheme requires the knowledge of the manifold structure, or equivalently the associated graph Laplacian operator \cite{poddar2015dynamic, poddar2019manifold, mohsin2019free}. The graph Laplacian operator, which can be viewed as the discrete approximation of the Laplace Beltrami operator, involves the similarity measurement among data points on the manifold. The shared intrinsic topology enables us to construct the graph Laplacian operator of the fingerprint manifold $\cal S$ using the low-dimensional parameter manifold $\cal M$. However, since fingerprints contaminated with aliasing artifacts and noise will decrease the accuracy of estimated tissue parameters, it is still challenging to directly estimate the manifold structure using acquired tissue fingerprints.
\begin{figure}[htbp]   
  \centering   
  \includegraphics[width=250pt]{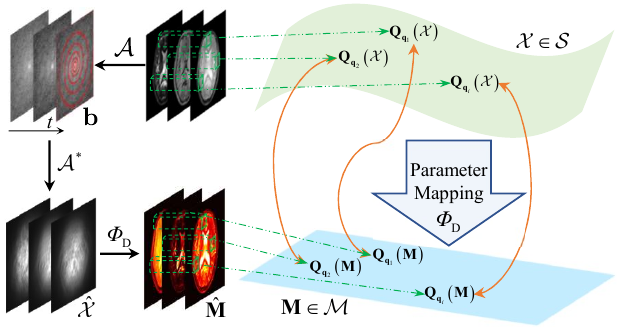}   
  \vspace{-22pt}
  \caption{Illustration of the proposed manifold structured data regularization scheme. The MRF data is divided into patches along the spatial dimension, and the similarity measurement among MRF data patches can be accurately obtained using the Euclidean distance between the corresponding patches in the parameter manifold.
  }   
  \label{manifold}   
  \vspace{-15pt}
\end{figure}

To address this issue, we propose to incorporate the manifold structured data priors in a spatially patch-wise manner, as illustrated in Fig.\ref{manifold}. Specifically, the MRF data is divided into overlapped spatial patches, and the similarity measurement among MRF data patches can be obtained using the Euclidean distance between the corresponding patches in the parameter manifold. The proposed scheme can exploit the non-linear and non-local redundancies in the MRF data, and thus improve the accuracy of similarity estimation and enhance robustness in undersampling and noisy scenarios.

The proposed manifold structured data regularization term ${\cal J}_{\rm MS}({\cal X})$ can be defined as:
\begin{equation}
  \label{manifold_regularization_1}
  {\cal J}_{\rm MS}({\cal X}) = \sum_{i} \sum_{j} w_{i,j} \lVert \textbf{\cal Q}_{\mathbf{q}_i}({\cal X}) - \textbf{\cal Q}_{\mathbf{q}_j}({\cal X}) \rVert_F^2 \quad i\neq j
\end{equation}
where $i$ and $j$ are indices of the data patches. $w_{i,j}$ represents the measured similarity weight between data patches. $\textbf{\cal Q}_{\mathbf{q}_k}(\cdot)$ denotes the operator that extracts data patch centered at the spatial location $\mathbf{q}_k$. $\textbf{\cal Q}_{\mathbf{q}_k}({\cal X}) \in {\mathbb C}^{p \times p \times L}$ denotes an extracted data patch with patch size $p\times p$ and stride $s = p|2$.

Instead of estimating similarity weight $w_{i,j}$ directly in high-dimensional space, we propose to compute the weight according to the Euclidean distance between the corresponding points on the low-dimensional parameter manifold ${\cal M}$, which can be modeled as:
\begin{equation}
  \label{weight}
  w_{i,j} = \exp \left( -\frac{\lVert \textbf{\cal Q}_{\mathbf{q}_i}({\hat{\mathbf{M}}}) - \textbf{\cal Q}_{\mathbf{q}_j}({\hat{\mathbf{M}}}) \rVert_F^2}{\sigma^2} \right)
\end{equation}
where $\sigma$ is a hyperparameter used to adjust the weight.

The regularization term in (\ref{manifold_regularization_1}) can be rewritten in matrix operation form by introducing the graph Laplacian operator $\mathbf{L}$:
\begin{equation}
  \label{manifold_regularization_2}
  {\cal J}_{\rm MS}({\cal X}) = {\rm Tr}\left( \textbf{\cal Q}({\cal X}) \mathbf{L} \textbf{\cal Q}({\cal X})^H \right)
\end{equation}
where ${\rm Tr}(\cdot)$ denotes the trace operator. $\textbf{\cal Q}({\cal X})$ is the operator that extracts data patches from $\cal X$ and arranges them into a Casorati matrix, where each column of the matrix corresponds to a vectorized data patch $\textbf{\cal Q}_{\mathbf{q}}({\cal X})$. The graph Laplacian operator $\mathbf{L}$ is defined as:
\begin{equation}
  \label{graph_laplacian}
  \mathbf{L} = \mathbf{D} - \mathbf{W}
\end{equation}
where $\mathbf{D}$ is a diagonal matrix with entries $\mathbf{D}_{i, i} = \sum_{j} w_{i,j}$, representing the degree of each node in the graph. $\mathbf{W}$ is the weighted adjacency matrix with entries defined by (\ref{weight}). For instance, in a three-dimensional space, the graph Laplacian operator is given by:
\begin{equation}
  \label{graph_laplacian_3d}
  \mathbf{L}  = \begin{bmatrix}
  w_{1,2}+w_{1,3} & -w_{1,2} & -w_{1,3} \\
  -w_{1,2} & w_{1,2}+w_{2,3} & -w_{2,3} \\
  -w_{1,3} & -w_{2,3} & w_{1,3}+w_{2,3}
  \end{bmatrix}
\end{equation}

The non-linear and non-local redundancies in the MRF data can be represented by the graph Laplacian operator, which is integrated into the proposed regularization term (\ref{manifold_regularization_2}) for improved MRF reconstruction.

\vspace{-5pt}
\subsection{MS-LLR Reconstruction Framework}
\label{sec:ms-llr}
Although our proposed manifold structured data regularization scheme is effective for capturing non-local and non-linear structure priors for MRF reconstruction, it may not be able to utilize the local correlations within each data patch. To overcome this limitation, we incorporate the locally low-rank regularization to further improve the reconstruction performance.

The proposed MRF reconstruction framework, termed MS-LLR, can be formulated as the following convex optimization problem:
\vspace{-5pt}
\begin{equation} 
  \label{opt_2}
  \min_{{\cal X}\in {\cal S} } \frac{1}{2}\lVert {\cal A} {\cal X} - \mathbf b \rVert_F^2 + \lambda_1 {\cal J}_{\rm MS}({\cal X}) + \lambda_2 {\cal J}_{\rm LLR}({\cal X}) \\
\end{equation}
where $\lambda_1$ and $\lambda_2$ are the corresponding tunable regularization parameters. ${\cal J}_{\rm MS}({\cal X})$ and ${\cal J}_{\rm LLR}({\cal X})$ are the manifold structured data regularization term and the locally low-rank regularization term, respectively. 

The adopted locally low-rank regularization can be expressed as:
\vspace{-5pt}
\begin{equation}
  \label{low_rank}
  {\cal J}_{\rm LLR}({\cal X}) = \sum_{i}\left\Vert \textbf{\cal Q}_{\mathbf{q}_i}({\cal X})\right\Vert_*
  \vspace{-5pt}
\end{equation}
where $\left\Vert \cdot \right\Vert_*$ denotes the nuclear norm and is used as the convex relaxation of the low-rank penalty to avoid NP-hard problems. The above locally low-rank regularization reuses the same patch extraction operator $\textbf{\cal Q}_{\mathbf{q}}$ as the manifold structured data regularization, which can reduce computation complexity. 

The manifold structured data regularization captures non-local and non-linear redundancies, while the locally low-rank regularization enhances the local correlations within each data patch. According to (\ref{manifold_regularization_2}) and (\ref{low_rank}), the proposed optimization problem can be specified as:
\vspace{-3pt}
\begin{equation}
  \label{optimization}
  \begin{aligned}
    \min_{{\cal X}\in {\cal S} } \frac{1}{2}\lVert {\cal A} {\cal X} - \mathbf b \rVert_F^2 
    &+  \lambda_1 {\rm Tr}\left(\textbf{\cal Q}({\cal X}) \mathbf{L} \textbf{\cal Q}({\cal X})^H\right) \\
    &+ \lambda_2 \sum_{i}\left\Vert \textbf{\cal Q}_{\mathbf{q}_i}({\cal X})\right\Vert_*
  \end{aligned}
  \vspace{-10pt}
\end{equation}

\vspace{-5pt}
\subsection{Optimization Algorithm}
We adopt the incremental subgradient-proximal method \cite{sra2012optimization} to solve (\ref{optimization}) iteratively. At the $n$-th iteration, it involves solving the following two subproblems:
\begin{align}
    &{\mathbf Z}^{n} \!=\! {\cal P_S}\left( {\cal X}^{n}\!-\!\mu\left[ {\cal A}^*({\cal A}{\cal X}^{n}\!-\!{\mathbf{b}})\!+\!\lambda_1 \textbf{\cal Q}^*(\textbf{\cal Q}({\cal X}^{n}) \mathbf{L}^{n})\right]\right)\label{incremental_proximal_01} \\
    &{\cal X}^{n+1} = \arg\min_{{\cal X}} \lambda_2 \sum_{i}\left\Vert \textbf{\cal Q}_{\mathbf{q}_i}({\cal X})\right\Vert_*+\frac{1}{2\mu}\lVert {\cal X}-{\mathbf{Z}}^{n} \rVert_F^2  
    \label{incremental_proximal_02}
\end{align}
where ${\cal A}^*$ and $\textbf{\cal Q}^*$ denote the adjoint operators of ${\cal A}$ and $\textbf{\cal Q}$, respectively. 
${\cal P_S}(\cdot)$ denotes the projection operator onto the fingerprint manifold ${\cal S}$, and $\mu>0$ is the step size. The projection operator can be defined as:
\vspace{-3pt}
\begin{equation}
  \label{projection_operator}
  {\cal P_S}(\mathbf{X}) = \mathbf{X}\mathbf{d}^\dagger\mathbf{d}
  \vspace{-3pt}
\end{equation}
where $\mathbf{d}$ is an orthonormal basis of the fingerprint manifold ${\cal S}$. It is worth noting that ${\cal S}$ is the cone on the Bloch response manifold $\cal B$, which shares the same orthonormal basis. Thus, the orthonormal basis $\mathbf{d}$ can be approximated by the precomputed dictionary $\mathbf{D}$ \cite{hu2021high}.

The subgradient iteration (\ref{incremental_proximal_01}) is used to update the auxiliary variable $\mathbf{Z}$, and the proximal iteration (\ref{incremental_proximal_02}) is used to update the MRF data ${\cal X}$. However, the proximal iteration cannot be solved analytically. Therefore, we adopt the variable splitting scheme and the alternating minimization scheme to efficiently solve it. By introducing a variable splitting scheme, the proximal iteration (\ref{incremental_proximal_02}) can be rewritten as the following constrained minimization problem:
\vspace{-5pt}
\begin{equation}
  \label{proximal_0}
  \min_{\cal X} \lambda_2 \sum_{i}\left\Vert \mathbf{P}_{\mathbf{q}_i}\right\Vert_*+\frac{1}{2\mu}\lVert {\cal X}-{\mathbf{Z}} \rVert_2^2 \quad s.t. \quad \mathbf{P}_{\mathbf{q}_i} = \textbf{\cal Q}_{\mathbf{q}_i}({\cal X}) 
  \vspace{-5pt}
\end{equation}
where $\mathbf{P}_{\mathbf{q}_i}$ denotes the auxiliary variable. The regularization penalties can be majorized using quadratic functions, and the constrained minimization problem can be rewritten as follows:
\vspace{-5pt}
\begin{equation}
  \label{proximal}
  \begin{aligned}
  \min_{\cal X} \lambda_2 \sum_{i}\left\Vert \mathbf{P}_{\mathbf{q}_i}\right\Vert_*&+\frac{1}{2\mu}\lVert {\cal X}-{\mathbf{Z}} \rVert_2^2\\
  &+ \frac{\lambda_2\beta}{2}\sum_{i}\lVert \mathbf{P}_{\mathbf{q}_i}-\textbf{\cal Q}_{\mathbf{q}_i}({\cal X}) \rVert_2^2
  \vspace{-5pt}
\end{aligned}
\end{equation}
where $\beta$ is the penalty parameter. The problem (\ref{proximal}) can be solved by the alternating minimization scheme, which alternates between solving the following two subproblems:
\begin{align}
  &\min_{\mathbf{P}_{\mathbf{q}_i}}  \sum_{i}\left(\lVert \mathbf{P}_{\mathbf{q}_i} \rVert_*+\frac{\beta}{2} \lVert \mathbf{P}_{\mathbf{q}_i}-\textbf{\cal Q}_{\mathbf{q}_i}({\cal X}) \rVert_2^2\right) \label{sub_proximal_1} \\
  &\min_{{\cal X}} \frac{1}{2\mu}\lVert {\cal X}-{\mathbf{Z}}^{(i)} \rVert_2^2+ \frac{\lambda_2\beta}{2}\sum_{i}\lVert \mathbf{P}_{\mathbf{q}_i}-\textbf{\cal Q}_{\mathbf{q}_i}({\cal X}) \rVert_2^2 \label{sub_proximal_2}
\end{align}

The subproblem (\ref{sub_proximal_1}) can be solved for each local data patch using the singular value thresholding (SVT) algorithm:
\begin{equation}
  \label{svt}
  \mathbf{P}_{\mathbf{q}_i}^{n+1} = \sum max(\sigma_i-\frac{1}{\beta},0)\mathbf{U}_i\mathbf{V}_i^H
\end{equation}
where $\textbf{\cal Q}_{\mathbf{q}_i}({\cal X}^{n}) =\sum \sigma_i\mathbf{U}_i\mathbf{V}_i^H$ is the singular value decomposition (SVD) of $\textbf{\cal Q}_{\mathbf{q}_i}({\cal X}^{(n)})$. The subproblem (\ref{sub_proximal_2}) is quadratic and can be solved analytically by:
\begin{equation}
  \label{sub_proximal_2_solution}
  {\cal X}^{n+1} = \frac{1}{1+\mu\lambda_2\beta}\left(\mathbf{Z}^{n}+\mu\lambda_2 \beta\textbf{\cal Q}^*(\mathbf{P}^{n+1}) \right)
\end{equation}
where $\mathbf{P}$ is a Casorati matrix with each column corresponding to a vectorized auxiliary variable $\mathbf{P}_{\mathbf{q}_i}$.

\begin{figure}[t]
  \vspace{-8pt}
\begin{algorithm}[H]
  \caption{Proposed MS-LLR Reconstruction Framework}
  \label{algorithm1}
  \begin{algorithmic}
      \STATE \textbf{Input:}
      \STATE Acquired $k$-space measurements $\mathbf{b}$;
      \STATE Predefined dictionary $\mathbf{D}$ and the corresponding parameter look-up table $\mathbf{LUT}$.
      \STATE \textbf{Output:}
      \STATE Magnetic parameter maps: $\mathbf{M}$.
      \STATE \textbf{Initialization:} 
      \STATE Hyper-Parameters: $\mu$, $\lambda_1$, $\lambda_2$, $\beta$, ${\cal P_S}(\cdot)$, $N_{max}$, 
      \STATE Iteration Init: ${\cal X}^0={\cal A}^*\mathbf{b}$, $\mathbf{M}^0 = \varPhi_{\mathbf{D}}({\cal X}^0)$, $\mathbf{L}^0$.
      \STATE \textbf{for} n=1 to $N_{max}$ \textbf{do}
      \STATE \quad Update of $\mathbf{Z}^n$ using (\ref{incremental_proximal_01});
      \STATE \quad Update of $\mathbf{P}^n$ using (\ref{svt});
      \STATE \quad Update of ${\cal X}^n$ using (\ref{sub_proximal_2_solution});
       
      \STATE \quad Compute the cost:
      \STATE \qquad $\textbf{\rm cost}(n)=\frac{1}{2}\lVert {\cal A} {\cal X} - \mathbf b \rVert_F  ^2 
      +  \lambda_1 {\rm Tr}\left(\textbf{\cal Q}({\cal X}) \mathbf{L} \textbf{\cal Q}({\cal X})^H\right)$ 
      \STATE \qquad \qquad \qquad \qquad\qquad\qquad\ $+ \lambda_2 \sum_{i}\left\Vert \textbf{\cal Q}_{\mathbf{q}_i}({\cal X})\right\Vert_*$
      \STATE \quad \textbf{if} $\textbf{\rm cost}(n-1)/\textbf{\rm cost}(n)-1<1e-5$
      \STATE \qquad \textbf{break}
      \STATE \quad Update $\mathbf{M}^{n+1} = \varPhi_{\mathbf{D}}({\cal X}^n)$;
      \STATE \quad Update $\mathbf{L}^{n+1}$ using (\ref{graph_laplacian});
      \STATE \quad Update $\lambda_1^n =\lambda_1^0 \times max(\mathbf{L}^n)$;
      \STATE \textbf{end}
      \STATE \textbf{Convergence Output:} $\hat{\cal X}$, $\hat{\mathbf{M}} =  \varPhi_{\mathbf{D}}(\hat{\cal X})$.
  \end{algorithmic}
\end{algorithm}
\vspace{-20pt}
\end{figure}
The implementation flow of the proposed MS-LLR scheme is shown in the pseudocode \textbf{Algorithm \ref{algorithm1}}.

\subsection{Implementation Details}
\label{sec:implementation}
Our proposed method was implemented on a Linux workstation with an Intel Xeon CPU and an Nvidia Quadro GV100 GPU. In particular, the size of the overlapped patch is set to $11\times 11$, and the step stride is set to $5$. For a finer balance of regularization terms, the penalty parameter $\lambda_1$ of the manifold regularization item is also dynamically adjusted with the maximum value of $\mathbf{L}$, i.e., $\lambda_1^n =\lambda_1^0 \times max(\mathbf{L}^n)$, and $\lambda_1^0$ is experimentally set to $0.1$. The other hyperparameters are experimentally set as $\mu=1$, $\lambda_2=0.1$, and $\beta={1}/{5}$. In addition, the algorithm is terminated till the relative change in the cost for successive iterations is less than a predefined tolerance value or until the maximum number of iterations is reached. The maximum number of iterations is set to $N_{max}=50$. We also adopt a GPU-accelerated NUFFT library \cite{muckley:20:tah} to accelerate the reconstruction performance of the proposed method in non-Cartesian sampling scenarios. 
\section{Experimental Results}
\label{sec:experiments}
In order to demonstrate the utility of the proposed method, we conducted experiments on the simulated and $\emph{in vivo}$ data for MRF reconstruction. We compare the performance of the proposed MS-LLR algorithm with several state-of-the-art methods, including the BLIP \cite{davies2014compressed}, MBIR \cite{zhao2016maximum}, FLOR \cite{mazor2018low}, and SL-SP \cite{hu2021high}. In addition, we also conduct ablation experiments to compare the reconstruction results using only locally low-rank constraints (LLR) to further verify the effectiveness of the proposed method. We have tuned the parameters for all the experiments to ensure optimal performance in each scenario.

\begin{figure}[htbp]
  \centering
  \includegraphics[width=250pt]{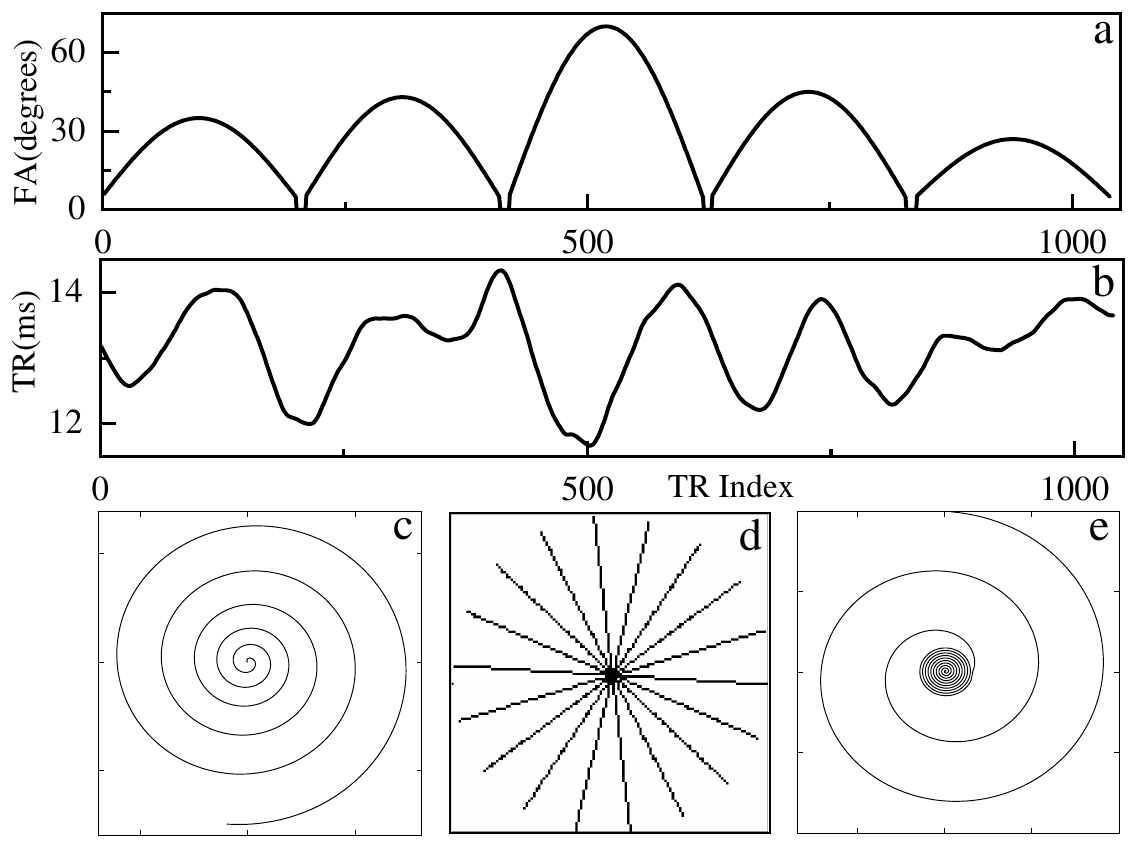}
  \vspace{-18pt}
  \caption{\textbf{a} and \textbf{b} are the flip angles and repetition time patterns that were used in the experiment. \textbf{c, d, e} shows the spiral undersampling trajectory, pseudo radial Cartesian sampling mask, and the variable density spiral undersampling trajectory, respectively, used in one repetition time in the experiments.}
  \label{fatr_mask}
  \vspace{-15pt}
\end{figure}

The same dictionary was used for all the experiments based on the following parameter discretization scheme: 1) $\rm T_1$ values were set within [100, 5000] ms, with an increment of 20 ms in the range of [100, 2000] ms, and an increment of 300 ms in the range of [2300, 5000] ms; 2) $\rm T_2$ values were set within [20, 1900] ms, with an increment of 5 ms in the range of [20, 100] ms, an increment of 10 ms in the range of [110, 200] ms, and an increment of 200 ms in the range of [300, 1900] ms. By omitting the combinations when the $\rm T_1$ values are less than the $\rm T_2$ values, the aforementioned setting results in a dictionary with the size of $3336\times L$.

\begin{figure*}[ht]
  \vspace{-5pt}
  \centering
  \includegraphics[width=420pt]{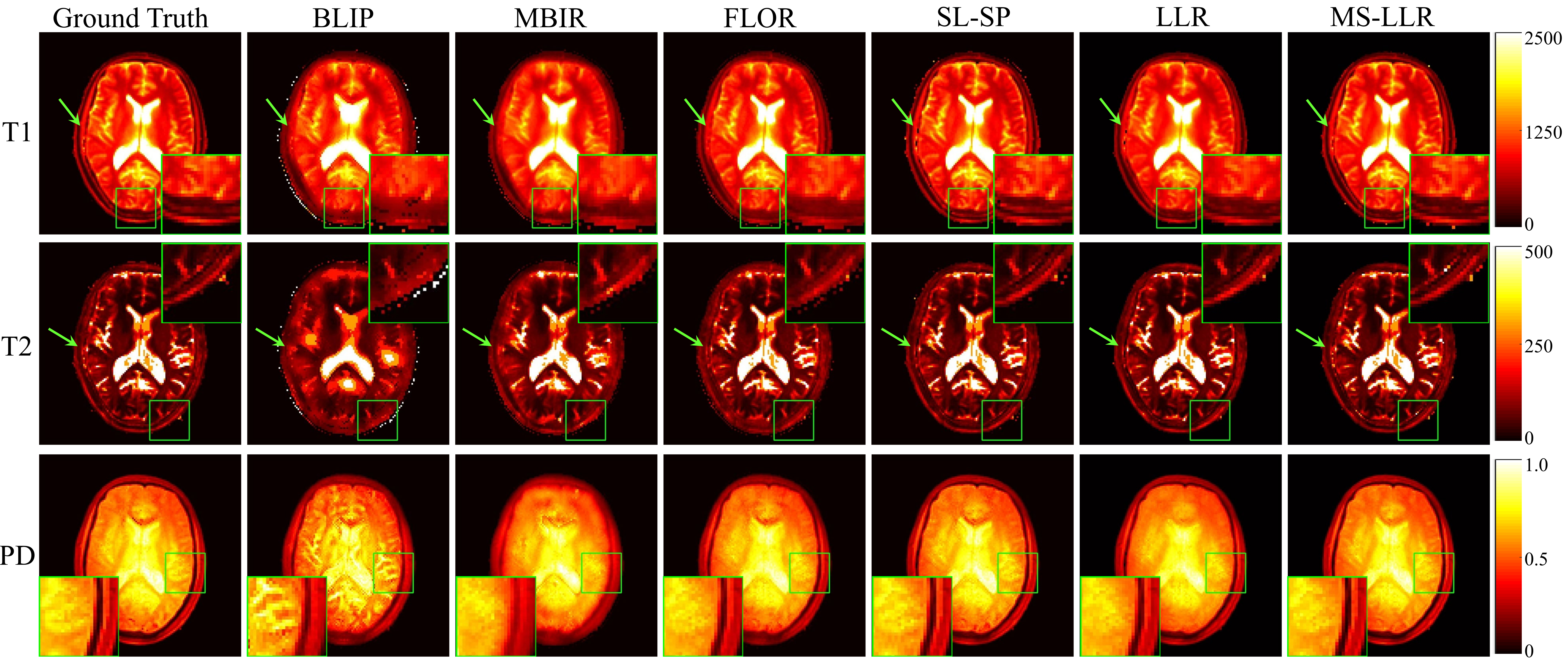}
  \vspace{-5pt}
  \caption{Reconstructed parameter maps of $\rm T_1$, $\rm T_2$, and PD. From the left to the right columns are the ground truth maps, estimated maps by BLIP, MBIR, FLOR, SL-SP, LLR, and the proposed MS-LLR method, with the acquisition length of $L=400$ using 5\% noiseless undersampled measurements.}
  \label{simu_400_free_vds}
\end{figure*}
\begin{figure*}
  \centering
  \vspace{-15pt}
  \hspace*{1.8cm}
  \includegraphics[width=360pt]{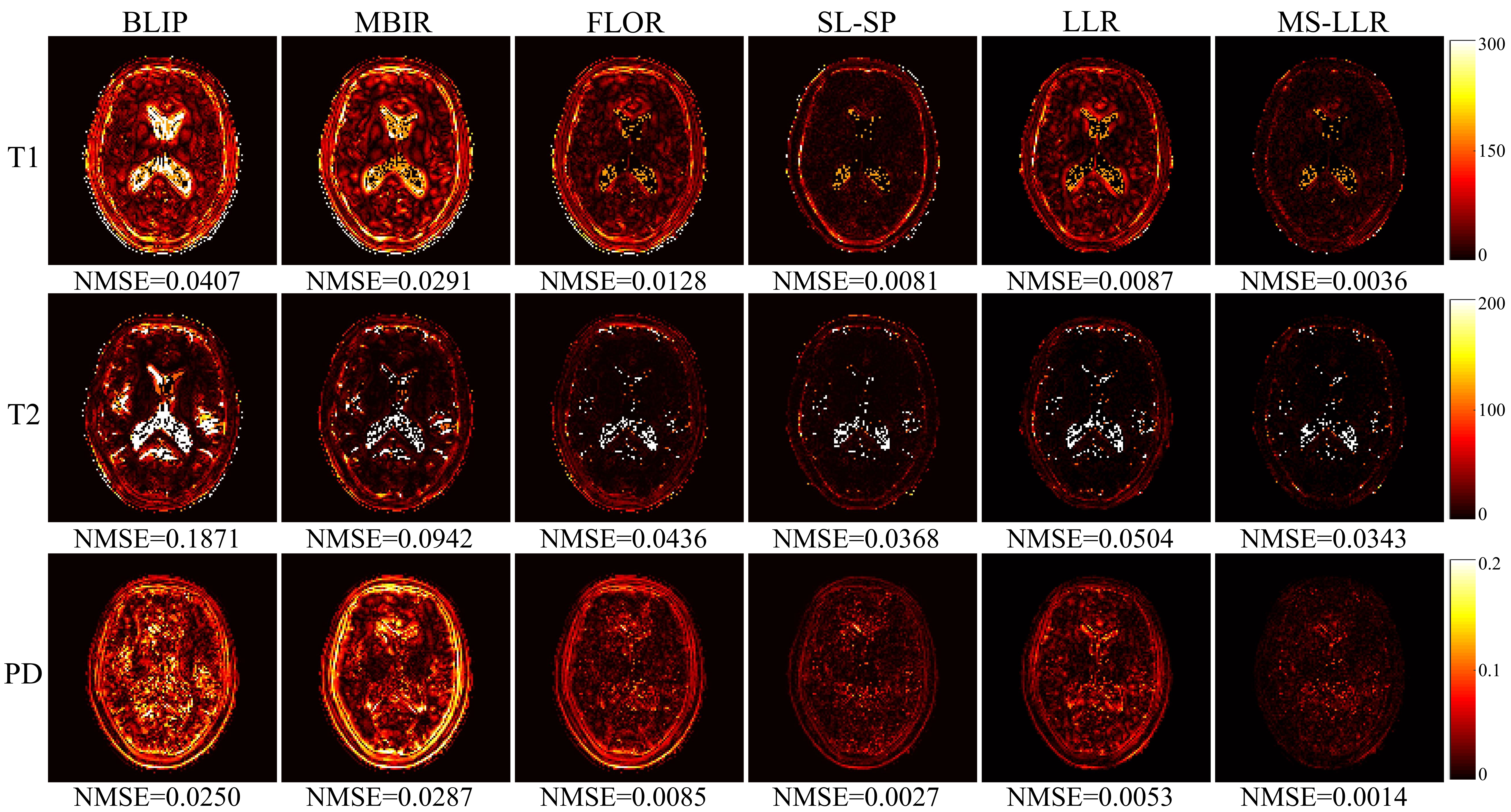}
  \vspace{-7pt}
  \caption{Error maps between the reconstructed parameter maps and the reference maps of $\rm T_1$, $\rm T_2$, and PD. From the left to the right columns are the error maps obtained by BLIP, MBIR, FLOR, SL-SP, LLR, and the proposed MS-LLR method, with the acquisition length of $L=400$ using 5\% noiseless undersampled measurements.}
 \label{simu_400_free_vds_error}
 \vspace{-10pt}
\end{figure*}

The fast imaging with steady-state precession (FISP) pulse sequence was used for all the experiments with FA and TR patterns as shown in Fig.\ref{fatr_mask} \textbf{a, b}. The echo time TE is fixed to $2.94$ ms. We used three different sampling trajectories in the experiments, as shown in Fig.\ref{fatr_mask} \textbf{c, d, e}.

We adopt two performance evaluation indexes to quantitatively evaluate the performance of the proposed method, including the signal-to-noise ratio (SNR) and the normalized mean square error (NMSE). Specifically, we use the signal-to-noise ratio (SNR) to measure the quality of the reconstructed MRF data, which can be defined as:
\begin{equation}
\label{eq:snr}
\rm{SNR}=-10\log_{10} \left(\frac{\lVert {\cal X} - \hat{{\cal X}} \rVert_\emph{F}^2}{\lVert {\cal X}\rVert_\emph{F}^2}\right)
\end{equation}
where ${\cal X}$ and $\hat {\cal X}$ are the ground truth and the reconstructed MRF data, respectively. The normalized mean square error (NMSE) is used to measure the quality of the reconstructed parameter maps $\hat{\mathbf{M}}$, which can be defined as:
\begin{equation}
  \label{nmse}
  {\rm NMSE} = \frac{\left\Vert \hat{m}_i - m_i \right\Vert^2}{\left\Vert m_i \right\Vert^2}
\end{equation}
where $m_i$ and $\hat{m}_i$ denote the ground truth and reconstructed parameters, respectively.
\subsection{Simulation Experiments}
\label{sec:simulation}
\begin{table*}[htbp]
  \vspace{-5pt}
  \centering
  \caption{The NMSEs of the reconstructed parameter maps by different methods using various acquisition lengths $L$.}
  \vspace{-3pt}
  \resizebox{480pt}{!}{
    \begin{tabular}{c|c|ccc|ccc|ccc|ccc}
    \toprule
          & L     & \multicolumn{3}{c|}{200} & \multicolumn{3}{c|}{300} & \multicolumn{3}{c|}{400} & \multicolumn{3}{c}{500} \\
          & Maps  & T1    & T2    & PD    & T1    & T2    & PD    & T1    & T2    & PD    & T1    & T2    & PD \\
    \midrule
    \multirow{6}[2]{*}{\begin{sideways}Noiseless\end{sideways}} & BLIP  & 0.0823  & 0.3241  & 0.0316  & 0.0614  & 0.2315  & 0.0255  & 0.0407  & 0.1871  & 0.0250  & 0.0320  & 0.1479  & 0.0248  \\
          & MBIR  & 0.0345  & 0.2891  & 0.0296  & 0.0327  & 0.1118  & 0.0289  & 0.0291  & 0.0942  & 0.0287  & 0.0276  & 0.0845  & 0.0254  \\
          & FLOR  & 0.0179  & 0.1136  & 0.0144  & 0.0131  & 0.0725  & 0.0093  & 0.0128  & 0.0436  & 0.0085  & 0.0102  & 0.0311  & 0.0067  \\
          & SL-SP & 0.0150  & 0.1076  & 0.0054  & 0.0081  & 0.0662  & 0.0029  & 0.0081  & 0.0368  & 0.0027  & 0.0075  & 0.0282  & 0.0022  \\
          & LLR   & 0.0159  & 0.1584  & 0.0121  & 0.0100  & 0.0870  & 0.0057  & 0.0087  & 0.0504  & 0.0053  & 0.0067  & 0.0319  & 0.0037  \\
          & MS-LLR & \textbf{0.0114 } & \textbf{0.1040 } & \textbf{0.0045 } & \textbf{0.0034 } & \textbf{0.0642 } & \textbf{0.0022 } & \textbf{0.0036 } & \textbf{0.0343 } & \textbf{0.0014 } & \textbf{0.0030 } & \textbf{0.0154 } & \textbf{0.0010 } \\
    \midrule
    \multirow{6}[2]{*}{\begin{sideways}Noisy\end{sideways}} & BLIP  & 0.0872  & 0.3789  & 0.0321  & 0.0682  & 0.2830  & 0.0311  & 0.0564  & 0.2080  & 0.0302  & 0.0453  & 0.1549  & 0.0291  \\
          & MBIR  & 0.0345  & 0.3490  & 0.0369  & 0.0344  & 0.2310  & 0.0353  & 0.0328  & 0.1524  & 0.0347  & 0.0311  & 0.1160  & 0.0318  \\
          & FLOR  & 0.0208  & 0.1676  & 0.0190  & 0.0143  & 0.1067  & 0.0131  & 0.0129  & 0.0680  & 0.0128  & 0.0104  & 0.0450  & 0.0101  \\
          & SL-SP & 0.0163  & 0.1417  & 0.0108  & 0.0112  & 0.0967  & 0.0105  & 0.0108  & 0.0667  & 0.0081  & 0.0102  & 0.0429  & 0.0075  \\
          & LLR   & 0.0186  & 0.1890  & 0.0132  & 0.0134  & 0.1117  & 0.0091  & 0.0120  & 0.0659  & 0.0088  & 0.0081  & 0.0400  & 0.0046  \\
          & MS-LLR & \textbf{0.0147 } & \textbf{0.1380 } & \textbf{0.0081 } & \textbf{0.0099 } & \textbf{0.0886 } & \textbf{0.0047 } & \textbf{0.0081 } & \textbf{0.0615 } & \textbf{0.0043 } & \textbf{0.0053 } & \textbf{0.0291 } & \textbf{0.0027 } \\
    \bottomrule
    \end{tabular}}
  \label{vds_nmse}
  \vspace{-15pt}
\end{table*}
In the simulation experiments, we evaluated the reconstruction performance of each algorithm using known quantitative parameters following the experimental setup in previous works \cite{mazor2018low,hu2021high}. The ground truth consisted of three known quantitative parameter matrices, $\rm T_1 \in [0, 4502]$, $\rm T_2 \in [0, 2547]$, and $\rm PD \in [0, 117]$, each with the size of $128 \times 128$. We used a variable density spiral trajectory to acquire 876 $k$-space coefficients in each frame, with the inner region size of 20 and FOV of 24 (see Fig.\ref{fatr_mask} \textbf{e}). The corresponding undersampling factor is $\sim 5\%$. We also conducted experiments by adding complex Gaussian white noise with $\sigma =0.5$ to the $k$-space data to simulate the noisy undersampled MRF measurements.

\begin{figure*}[t]
  \vspace{-5pt}
  \centering
  \includegraphics[width=420pt]{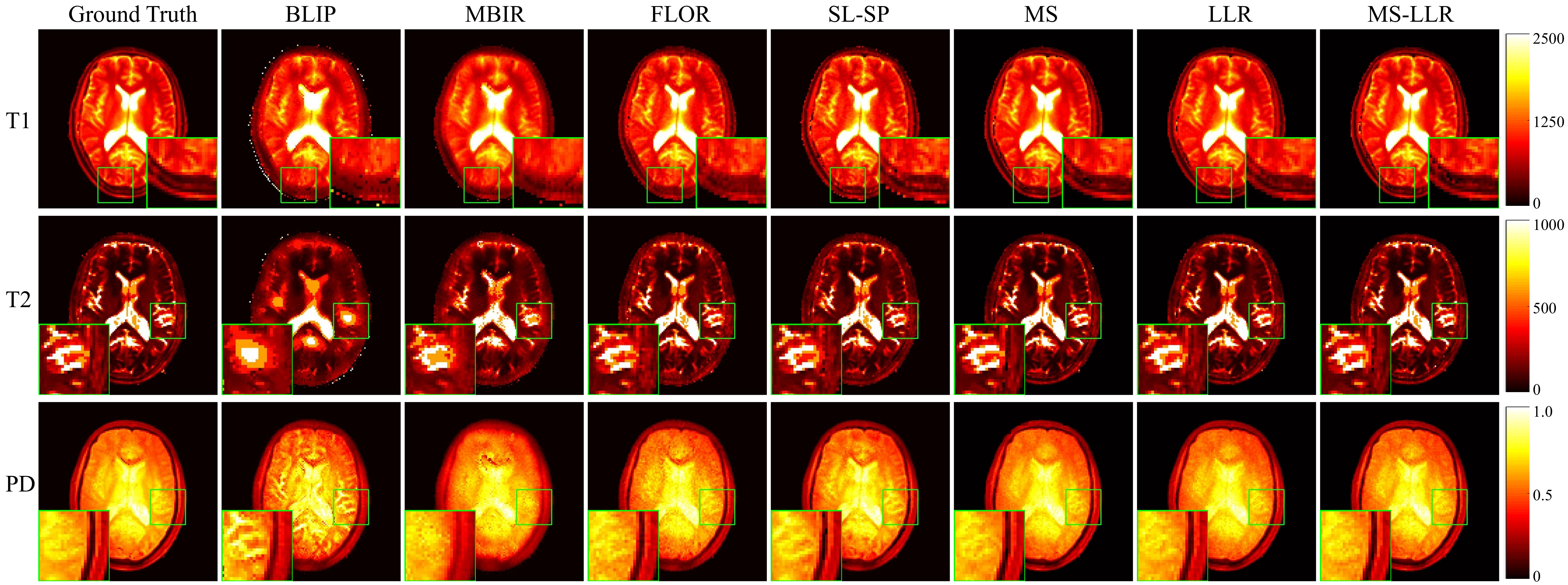}
  \vspace{-7pt}
  \caption{Reconstructed parameter maps of $\rm T_1$, $\rm T_2$, and PD. From the left to the right columns are the ground truth maps, estimated maps by BLIP, MBIR, FLOR, SL-SP, LLR, and the proposed MS-LLR method, with the acquisition length of $L=500$ using 5\% noisy undersampled measurements.}
  \label{simu_500_noisy}
\end{figure*}
\begin{figure*}
  \centering
  \vspace{-15pt}
  \hspace*{1.8cm}
  \includegraphics[width=360pt]{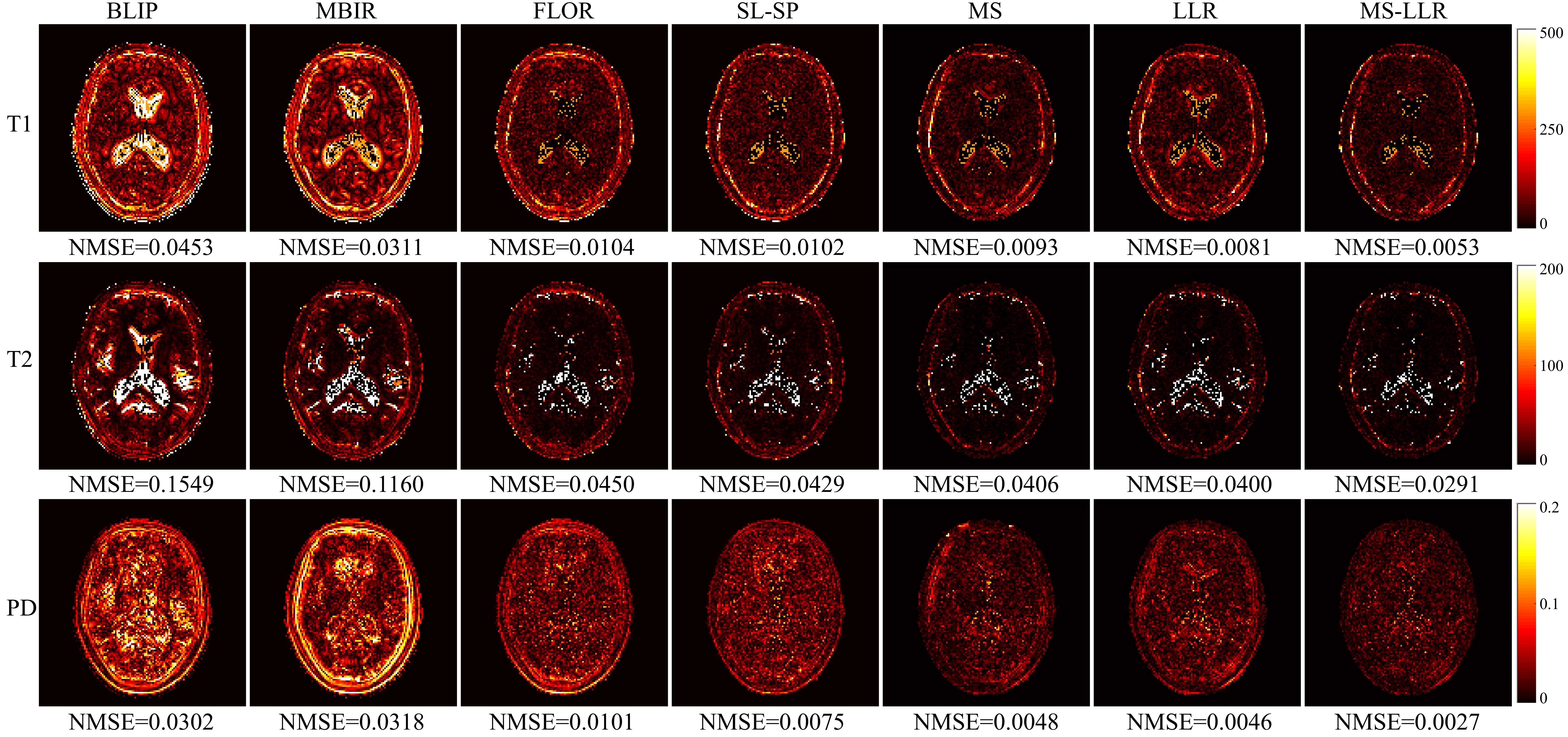}
  \vspace{-7pt}
  \caption{Error maps between the reconstructed parameter maps and the reference maps of $\rm T_1$, $\rm T_2$, and PD. From the left to the right columns are the error maps obtained by BLIP, MBIR, FLOR, SL-SP, LLR, and the proposed MS-LLR method, with the acquisition length of $L=500$ using 5\% noisy undersampled measurements.}
 \label{simu_500_noisy_error}
 \vspace{-15pt}
\end{figure*}

Fig.\ref{simu_400_free_vds} showed the reconstruction results obtained using 5\% noiseless undersampled data with an acquisition length of 400. The first column displayed the ground truth maps, while the 2nd through 7th columns corresponded to the reconstructed parameter maps of $\rm T_1$, $\rm T_2$, and PD obtained using BLIP, MBIR, FLOR, SL-SP, LLR, and the proposed MS-LLR method, respectively. The PD maps were normalized in the range of $[0,1]$ for simplicity. Fig.\ref{simu_400_free_vds_error} showed the corresponding error map to more clearly highlight the quality of the reconstructed parameter maps. We observed that due to the high undersampling factor, the BLIP and MBIR methods suffer from obvious blurring undersampling artifacts. By exploiting the correlation priors (low-rank and structured low-rank) of MRF data, the FLOR and SL-SP methods can provide improved reconstruction results with acceptable parameter map details. However, by combining the latent manifold structure priors and the locally low-rank constraints, the proposed MS-LLR method presented an optimal performance by providing the reconstructed maps with the highest accuracy. Moreover, the LLR method was only slightly better than the low-rank-based methods (FLOR), confirming that the proposed manifold structure prior can effectively improve the reconstruction performance. Fig.\ref{simu_500_noisy} and Fig.\ref{simu_500_noisy_error} showed the reconstruction results using 5\% noisy undersampled data with an acquisition length of 500. It was shown that the MS-LLR method performs the best in recovering tissue parameter maps, which was consistent with the noiseless scenario.

\begin{figure}[htbp]
  \vspace{-8pt}
  \centering
  \includegraphics[width=\columnwidth]{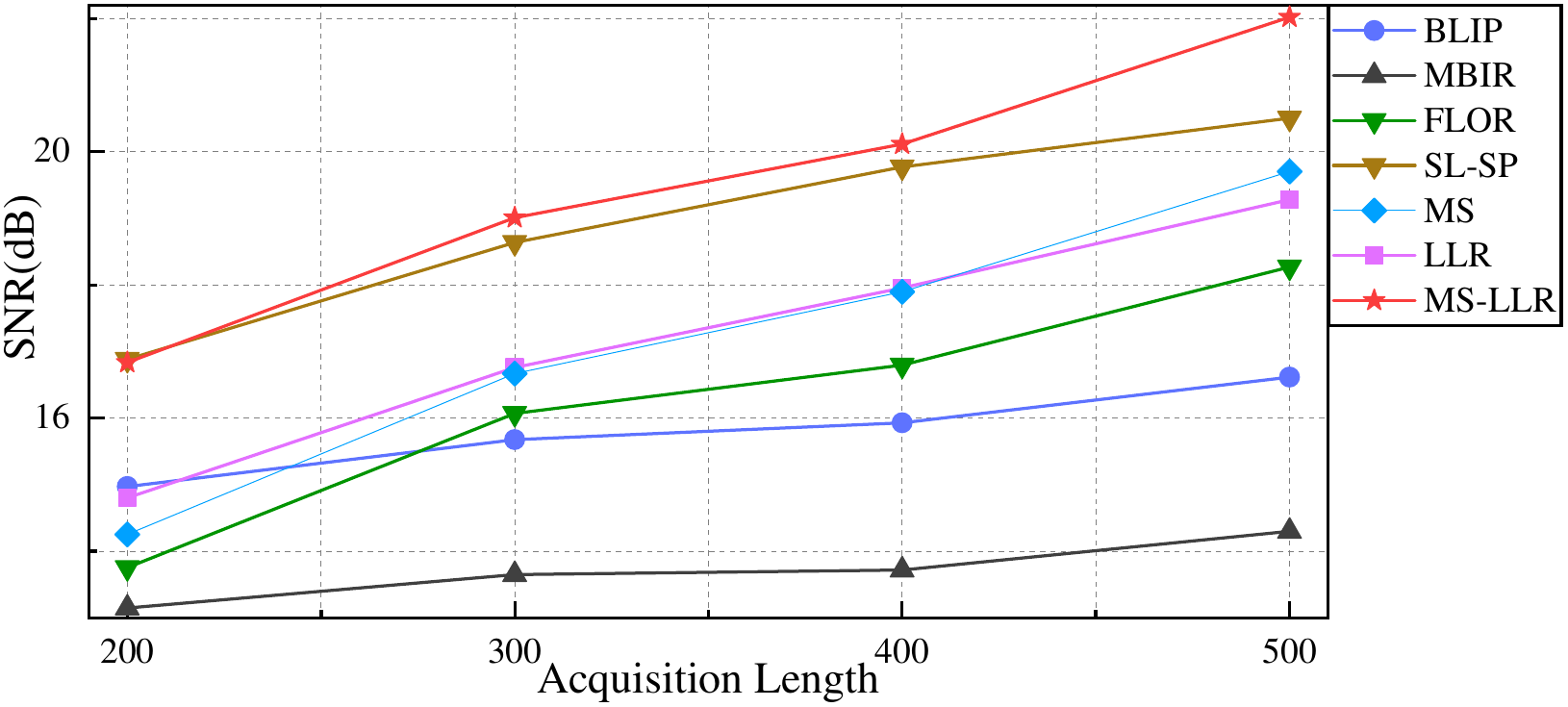}
  \vspace{-18pt}
  \caption{The SNRs (dB) of the space-time matrix $\cal X$ reconstructed by different methods using various acquisition length $L$. \textbf{a} shows the SNRs with different acquisition lengths using 5\% noiseless undersampled data. \textbf{b} shows the SNRs with different acquisition lengths using 5\% noisy undersampled data.}
  \label{vds_snr}
  \vspace{-12pt}
\end{figure}

We also studied the effect of different acquisition lengths on the reconstruction performance of each algorithm. The signal-to-noise ratio (SNR) of the reconstructed space-time matrices was plotted according to different acquisition lengths for all the methods under comparison using both noiseless and noisy undersampled data, as shown in Fig.\ref{vds_snr}. The NMSEs of the results using different acquisition lengths were reported in Table.\ref{vds_nmse}. The experimental results indicated that the reconstruction performance of all algorithms improved as the length of the acquired data increased, while the proposed MS-LLR method consistently provided optimal reconstruction performance. Notably, the MS-LLR method showed an improvement of approximately 3 dB over the state-of-the-art algorithms.

\begin{figure*}[htbp]
  \centering
  \vspace{-5pt}
  \includegraphics[width=480pt]{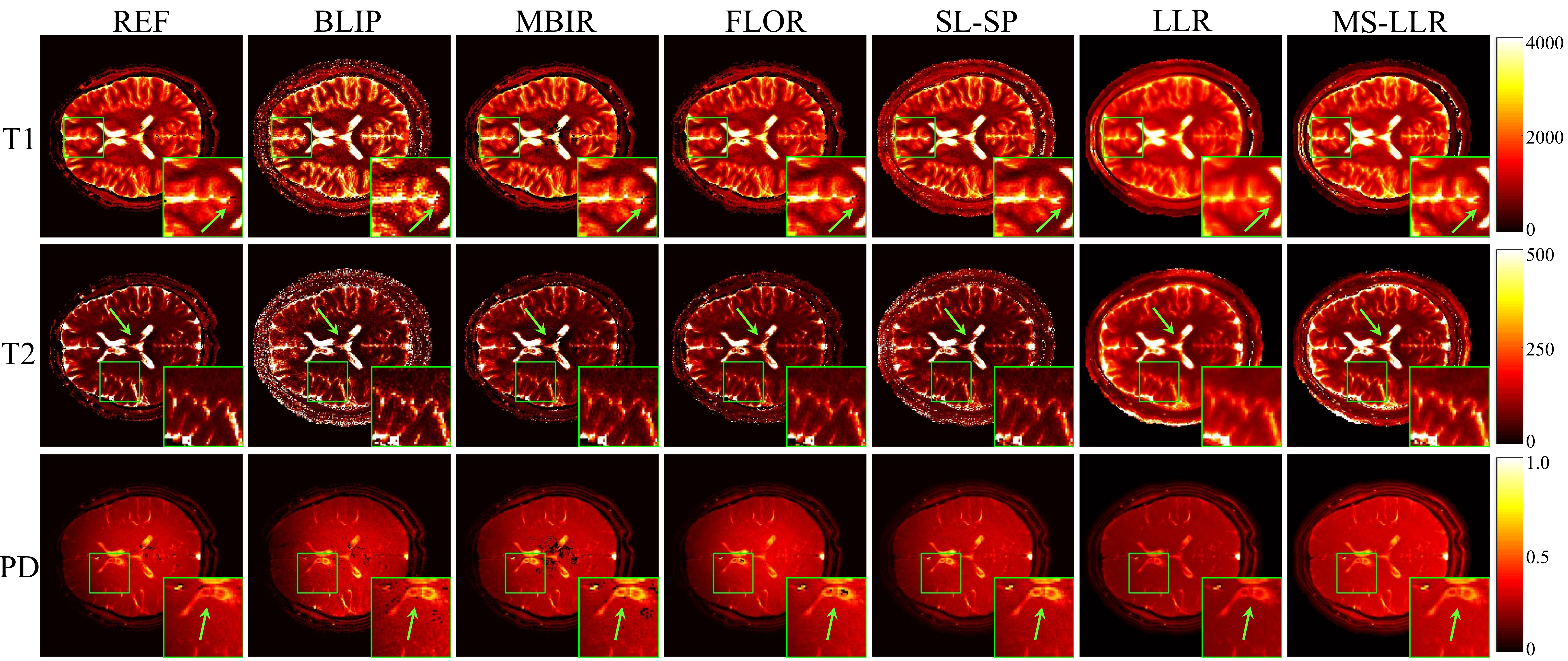}
  \vspace{-8pt}
  \caption{Reconstructed parameter maps of $\rm T_1$, $\rm T_2$, and $\rm PD$. From the left to the right columns are the reference maps (reconstructed using the original MRF method with the acquisition length of 1000), estimated maps with the acquisition length of $L=500$ by BLIP, MBIR, FLOR, SL-SP, LLR, and the proposed MS-LLR method, using 6\% undersampled Fourier measurements of the \emph{in vivo} data.}
  \label{invivo}
  \vspace{-10pt}
\end{figure*}
\subsection{In Vivo Experiments}
The \emph{in vivo} data used in this section were acquired on a 3T Siemens Prisma scanner from one healthy human subject using the FISP sequence with a 16-channel head coil. The acquisition utilized 36 spiral trajectories (see Fig.\ref{fatr_mask} \textbf{c}) to acquire 2880 samples per frame, resulting in an undersampling ratio of $\sim 6\%$. The imaging parameters used were FOV of $220\times 220 $ mm$^2$ and slice thickness of 5 mm. As fully undersampled ground truth data was not provided for the \emph{in vivo} experiment, we followed the approach used in \cite{hu2021high} and used the parameter maps estimated from data with an acquisition length of 1000 by the original MRF method as the reference. The parameter maps reconstructed by different methods with an acquisition length of 500 were shown in Fig.\ref{invivo}. It was observed that the BLIP and MBIR methods could recover sharper image details but at the cost of introducing noise-like isolated points. The FLOR and SL-SP methods showed improved reconstruction results by providing more accurate parameter estimates. The proposed MS-LLR algorithm provided the best reconstruction performance with the clearest tissue details in the reconstructed parameter maps.
\section{Discussion}
\label{sec:discussion}
In this section, we discuss the parameter settings and analyze the influence of the patch size and the sampling patterns on the performance of the proposed method. We also report the computational cost of different MRF reconstruction methods. 
\subsubsection{Parameter Settings}
\label{sec:para}
\begin{figure}[htbp]
  \centering
  \vspace{-10pt}
  \setlength{\abovecaptionskip}{-2pt}   
  \includegraphics[width=160pt]{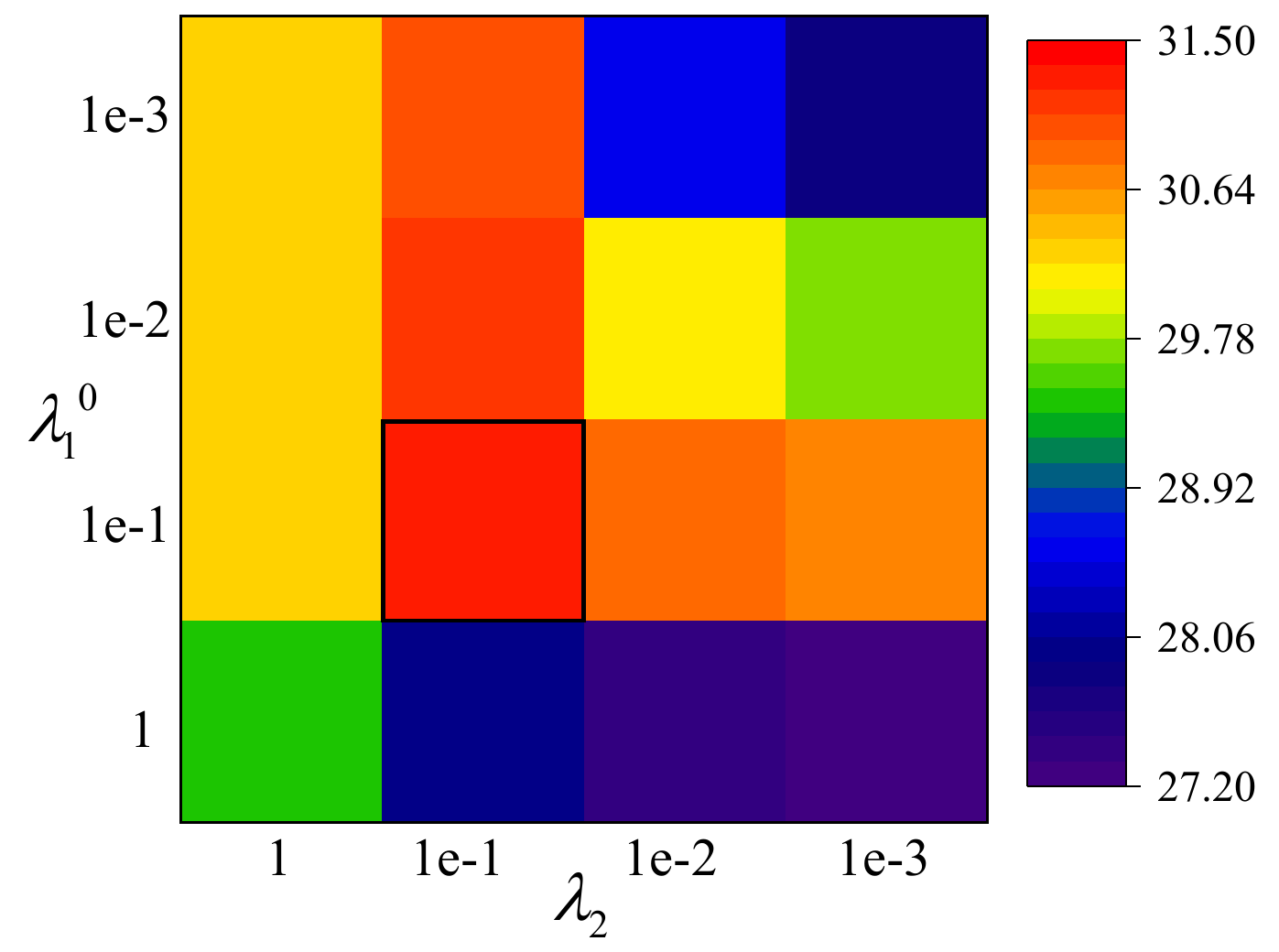}
  \caption{SNR values of the reconstructed MRF data $\cal X$ as function of the parameters $\lambda_1^0$ and $\lambda_2$. The region in bold indicates the maximum SNR values, where the optimal parameters are chosen.}
  \label{lambda}
  \vspace{-10pt}
\end{figure}

To investigate the optimal selection of the parameters $\lambda_1^0$ and $\lambda_2$ for achieving the best algorithm performance, we conducted experiments using pseudo-radial Cartesian trajectories (Fig.\ref{fatr_mask} \textbf{d}) with $L = 500$. The reconstruction results for various parameter combinations were presented in Fig.\ref{lambda}. The optimal parameter setting,  indicated by the bold area, was found to be $\lambda_1^0=0.1$ and $\lambda_2=0.1$.
\subsubsection{Patch Size}
To investigate the influence of the patch size on the reconstruction performance, we conducted experiments using pseudo-radial Cartesian trajectories (Fig.\ref{fatr_mask} \textbf{d}) with $L = 500$. The reconstruction results for various patch sizes were presented in Table.\ref{patch_size}. We can observe that the performance of the algorithm was not sensitive to the patch size, and the optimal patch size was $11\times 11$.
\begin{table}[htbp]
  \centering
  \vspace{-10pt}
  \caption{The NMSEs of the reconstructed MRF data $\cal X$ using different patch sizes.}
  \resizebox{\columnwidth}{!}{
    \begin{tabular}{c|cccccc}
    \toprule
    Patch Size & $5\times 5$     & $7\times 7$     & $9\times 9$    & $11\times 11 $  & $13\times 13$   & $15\times 15$\\
    \midrule
    T1    & 0.00413  & 0.00404  & 0.00442  & \textbf{0.00313 } & 0.00668  & 0.01026  \\
    T2    & 0.02071  & 0.02453  & 0.02454  & \textbf{0.01377 } & 0.02305  & 0.01973  \\
    PD    & 0.00099  & 0.00090  & 0.00089  & \textbf{0.00088 } & 0.00089  & 0.00090  \\
    \bottomrule
    \end{tabular}}
  \label{patch_size}
  \vspace{-8pt}
\end{table}
\subsubsection{Sampling Patterns}
To evaluate the effectiveness of the proposed model with different undersampling patterns, we carried out experiments to reconstruct the parameter maps of the simulated data using $\sim 5\%$ undersampled measurements at the acquisition length of 500 with three undersampling patterns, as illustrated in Fig.\ref{fatr_mask} \textbf{c-e}. The NMSEs of the reconstructed parameter maps are reported in Table.\ref{sampling_pattern}. The results indicate that the proposed method can consistently provide optimal results using different sampling trajectories.
\begin{table}[htbp]
  \centering
  \vspace{-10pt}
  \caption{The NMSEs of the reconstructed parameter maps using different methods under various sampling patterns.}
  \resizebox{\columnwidth}{!}{
    \begin{tabular}{c|ccc|ccc|ccc}
    \toprule
          & \multicolumn{3}{c|}{Spiral} & \multicolumn{3}{c|}{Vds-spiral} & \multicolumn{3}{c}{Radial} \\
    \midrule
          & T1    & T2    & PD    & T1    & T2    & PD    & T1    & T2    & PD \\
    \midrule
    BLIP  & 0.0963  & 0.4682  & 0.0169  & 0.0320  & 0.1479  & 0.0248  & 0.0218  & 0.0983  & 0.0130  \\
    MBIR  & 0.0582  & 0.3763  & 0.0404  & 0.0276  & 0.0845  & 0.0254  & 0.0146  & 0.0809  & 0.0065  \\
    FLOR  & 0.0175  & 0.1024  & 0.0091  & 0.0102  & 0.0311  & 0.0067  & 0.0051  & 0.0274  & 0.0013  \\
    SL-SP & 0.0094  & 0.0545  & 0.0039  & 0.0075  & 0.0282  & 0.0022  & \textbf{0.0027 } & 0.0246  & 0.0010  \\
    LLR   & 0.0166  & 0.0864  & 0.0113  & 0.0067  & 0.0319  & 0.0037  & 0.0035  & 0.0285  & 0.0013  \\
    MS-LLR & \textbf{0.0131 } & \textbf{0.0534 } & \textbf{0.0048 } & \textbf{0.0030 } & \textbf{0.0154 } & \textbf{0.0010 } & 0.0031  & \textbf{0.0138 } & \textbf{0.0009 } \\
    \bottomrule
    \end{tabular}}
  \label{sampling_pattern}
  \vspace{-7pt}
\end{table}
\subsubsection{Computational cost}
The proposed method leveraged both the manifold structure prior and local low-rank prior with a shared block extraction operator, which not only strengthens the utilization of the data structure prior but also reduces the complexity of the algorithm. Moreover, a GPU-accelerated NUFFT library \cite{muckley:20:tah} was employed to enhance the computational efficiency of our method in non-Cartesian sampling scenarios. To evaluate the computational efficiency of the proposed method, we conducted experiments using variable density spiral trajectory (Fig.\ref{fatr_mask} \textbf{e}) with a data acquisition length of 500. The computational time of various methods was reported in Table.\ref{time}, demonstrating that our proposed method achieves significant acceleration under non-Cartesian sampling patterns compared to state-of-the-art methods.
\begin{table}[htbp]
  \vspace{-10pt}
  \centering
  \caption{The computational time of different methods with non-Cartesian sampling patterns (in minutes).}
  \resizebox{240pt}{!}{
    \begin{tabular}{ccccccc}
    \toprule
    Methods & BLIP & MBIR & FLOR & SL-SP & LLR & MS-LLR \\
    \midrule
    Time  & 45.78 & 129.05 & 50.13 & 59.55 & 4.21  & 11.31 \\
    \bottomrule
    \end{tabular}}
  \label{time}
  \vspace{-10pt}
\end{table}

\section{Conclusion}
\label{sec:conclusions}
In this paper, we proposed a novel MRF reconstruction method combining manifold structured data priors and locally low-rank constraints. We revealed that the fingerprint manifold shared the same intrinsic topology as the parameter manifold, which enabled accurate estimating of the fingerprint manifold structure by leveraging the parameter manifold. By mining non-local and non-linear redundancies, as well as utilizing local data correlations, the proposed method demonstrated significant improvement over state-of-the-art methods in terms of reconstruction performance and efficiency. This scheme was efficient and robust to noise and undersampling, and can be easily extended to other MRF reconstruction methods. Furthermore, by utilizing a GPU-accelerated NUFFT library, the proposed method can achieve fast reconstruction with non-Cartesian sampling patterns. Future research can focus on further improving computational efficiency and generalization in more complex scenarios.

\bibliographystyle{IEEEtran}
\bibliography{refs.bib}

\end{document}